\documentclass[aps,prd,reprint,twocolumn,superscriptaddress,showpacs]{revtex4-1}
\usepackage{amsfonts}
\usepackage{mathrsfs}
\usepackage{amsmath}
\usepackage{color}
\usepackage{natbib}
\usepackage{textcomp}
\usepackage{graphicx}
\usepackage{bm}
\usepackage{amssymb}
\usepackage{xspace}
\usepackage{epstopdf}
\usepackage{dcolumn}
\usepackage{longtable}
\usepackage{multirow}
\usepackage{float}
\usepackage{bm}

\makeatletter
\newcommand{\Rmnum}[1]{\expandafter\@slowromancap\romannumeral #1@}
\makeatother
\begin{document}

\title{Kagome quantum anomalous Hall effect with high Chern number and large band gap}

\author{Zhen Zhang}
\affiliation{School of Physical Sciences, University of Chinese Academy of Sciences, Beijng 100049, China}

\author{Jing-Yang You}
\affiliation{Kavli Institute for Theoretical Sciences, and CAS Center for Excellence in Topological Quantum Computation, University of Chinese Academy of Sciences, Beijng 100190, China}

\author{Xing-Yu Ma}
\affiliation{School of Physical Sciences, University of Chinese Academy of Sciences, Beijng 100049, China}

 \author{Bo Gu}
 \email{gubo@ucas.ac.cn}
 \affiliation{Kavli Institute for Theoretical Sciences, and CAS Center for Excellence in Topological Quantum Computation, University of Chinese Academy of Sciences, Beijng 100190, China}
\affiliation{Physical Science Laboratory, Huairou National Comprehensive Science Center, Beijing 101400, China}

\author{Gang Su}
\email{gsu@ucas.ac.cn}
\affiliation{Kavli Institute for Theoretical Sciences, and CAS Center for Excellence in Topological Quantum Computation, University of Chinese Academy of Sciences, Beijng 100190, China}
\affiliation{Physical Science Laboratory, Huairou National Comprehensive Science Center, Beijing 101400, China}
\affiliation{School of Physical Sciences, University of Chinese Academy of Sciences, Beijng 100049, China}

\begin{abstract}
Due to the potential applications in the low-power-consumption spintronic devices, the quantum anomalous Hall effect (QAHE) has attracted tremendous attention in past decades. However, up to now, QAHE was only observed experimentally in topological insulators with Chern numbers $C$= 1 and 2 at very low temperatures. Here, we propose three novel two-dimensional stable kagome ferromagnets Co$_3$Pb$_3$S$_2$, Co$_3$Pb$_3$Se$_2$ and Co$_3$Sn$_3$Se$_2$ that can realize QAHE with high Chern number of $\lvert$$C$$\rvert$=3. Monolayers Co$_3$Pb$_3$S$_2$, Co$_3$Pb$_3$Se$_2$ and Co$_3$Sn$_3$Se$_2$ possess the large band gap of 70, 77 and 63 meV with Curie temperature $T_C$ of 51, 42 and 46 K, respectively. By constructing a heterostructure Co$_3$Sn$_3$Se$_2$/MoS$_2$, its $T_C$ is enhanced to 60 K and the band gap keeps about 60 meV due to the tensile strain of 2\% at the interface. For the bilayer compound Co$_6$Sn$_5$Se$_4$, it becomes a half-metal, with a relatively flat plateau in its anomalous Hall conductivity corresponding to $\lvert$$C$$\rvert$ = 3 near the Fermi level. Our results provide new topological nontrivial systems of  kagome ferromagnetic monolayers and heterostructrues possessing QAHE with high Chern number $\lvert$$C$$\rvert$ = 3 and large band gaps.

\end{abstract}
\maketitle


\section{\uppercase\expandafter{\romannumeral1}. Introduction}
The quantum anomalous Hall effect (QAHE) was predicted by Haldane in 1988 using a honeycomb lattice model~\cite{Haldane1988}. The QAHE is characterized by the nonzero Chern numbers with the quantized Hall conductance. Owing to the existence of dissipationless chiral edge states, QAHE can be used to design low-power-consumption spintronic devices. Up to now, many efforts have been done to search for QAHE materials with large band gap and high Curie temperature ($T_C$), including adsorbing diluted 3$d$, 4$d$ and 5$d$ transition atoms on graphene~\cite{Ding2011,Acosta2014,Zhang2012,Hu2015}; using surface functionalization on the buckled honeycomb lattice silicene, germanene and stanlene~\cite{Huang2014,Wu2014}; constructing heterostructure by depositing atomic layers of elements on a ferromagnetic insulator~\cite{Garrity2013,Qiao2014,Wang2015a,Zhang2015,Zhang2018,Zou2020}; applying pure ferromagnetic semiconductor compounds MX$_3$ without external fields or additional doping~\cite{Sheng2017,Wang2018,You2019a,You2019b}. Despite a lot of theoretical predictions, the observation of QAHE in experiment is rare. The first observed QAHE with quantized Hall conductance $e^2/h$ was in Cr-doped (Bi, Sb)$_2$Te$_3$ thin films at a much low temperature of 30mK~\cite{Chang2013}. Later, the QAHE in V-doped and Cr-and-V co-doped (Bi, Sb)$_2$Te$_3$ thin film were observed successively at about 25 and 300 mK, respectively~\cite{Chang2015,Ou2017}.
Recently, the quantized anomalous Hall conductance (AHC) plateau $\sigma_{xy} = e^2/h$ was observed in MnBi$_2$Te$_4$ thin flakes~\cite{Deng2020,Liu2020}, moreover, a high-Chern-number QAHE with $\sigma_{xy} = 2e^2/h$ was also obtained in ten-layer MnBi$_2$Te$_4$ device at about 13 K ~\cite{Ge2020}. However, there have been sparse theoretical reports of the QAHE with higher Chern number such as $C$=3~\cite{Miert2014,Cai2015,Song2016,Kong2018}.

Inspired by the Haldane model, in the past decades, most of the QAHE were predicted based on a honeycomb lattice. Nevertheless, kagome lattice with out-of-plane magnetization is also an important platform for investigating the QAHE~\cite{Ohgushi2000,Zhang2011,Xu2015,Yin2020}. In particular, layered magnetic kagome lattice Co$_3$Sn$_2$S$_2$ was recently reported to be a Weyl semimetal with a large intrinsic AHC~\cite{Liu2018}. Because of the successful synthesis of the bulk Co$_3$Sn$_2$S$_2$, monolayer Co$_3$Sn$_3$S$_2$ was studied theoretically, and was found to be a Chern insulator with $C$ = 3~\cite{Muechler2020}.

In this work, inspired by recent studies on Co$_3$Sn$_3$S$_2$, we systematically investigate monolayers Co$_3$X$_3$Y$_2$ (X = C,Si,Ge,Sn,Pb; Y =O,S,Se,Te,Po) based on the first principles calculations. Our results show that monolayers Co$_3$Pb$_3$S$_2$, Co$_3$Pb$_3$Se$_2$ and Co$_3$Sn$_3$Se$_2$ are stable. According to the results of anomalous Hall conductivity $\sigma_{xy}$ and Chiral edge states, a high Chern number $\lvert$$C$$\rvert$ = 3 was obtained in these three compounds. Furthermore, we find that the band gap and $T_C$ are all sensitive to the applied strain. For Co$_3$Sn$_3$Se$_2$ monolayer, its band gap can be decreased to zero with compressive strain of -3\%, and its $T_C$ can be increased to 65K with tensile strain of 4\%. $T_C$ = 60 K and tensile strain of 2\% can be obtained by constructing a heterostructure Co$_3$Sn$_3$Se$_2$/MoS$_2$. We have also explored the topological properties of bilayer Co$_3$Sn$_3$Se$_2$, and found a Weyl node on the $\Gamma\rightarrow$ M path near the Fermi level. Although there is no global gap, we find a relatively flat plateau in AHC corresponding to $\lvert$$C$$\rvert$ = 3 near the Fermi level.

\begin{figure*}[!!!hbt]
  \centering
  \includegraphics[scale=0.9,angle=0]{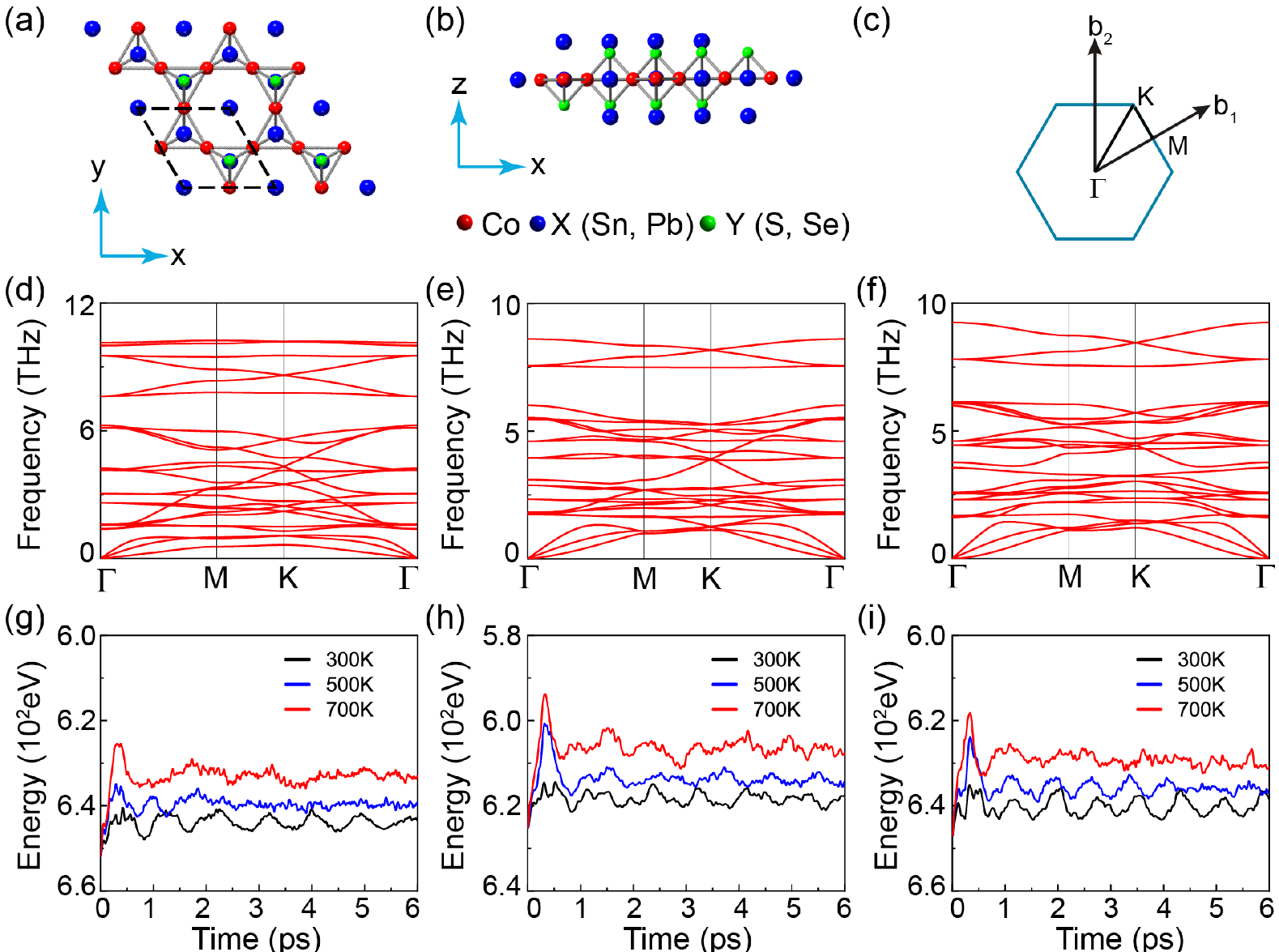}\\
  \caption{(a) Top and (b) side views of monolayer Co$_3$X$_3$Y$_2$. (c) First Brillouin zone. Phonon spectra of (d) Co$_3$Pb$_3$S$_2$, (e) Co$_3$Pb$_3$Se$_2$ and (f) Co$_3$Sn$_3$Se$_2$ monolayers. Molecular dynamics simulations for (g) Co$_3$Pb$_3$S$_2$, (h) Co$_3$Pb$_3$Se$_2$ and (i) Co$_3$Sn$_3$Se$_2$ at different temperatures for 6ps with a time step of 3fs.}\label{fig1}
\end{figure*}

\section{\uppercase\expandafter{\romannumeral2}. Computational Methods}
In our studies, the first-principles calculations were performed using the projector augmented wave (PAW) method~\cite{Bloechl1994} based on the density functional theory (DFT) as implemented in the Vienna ab initio simulation package (VASP)~\cite{Kresse1993,Kresse1996}. The electron exchange-correlation functional is described by the generalized gradient approximation (GGA) in the form proposed by Perdew, Burke, and Ernzerhof (PBE) \cite{Perdew1996}. A 20 {\AA} vacuum space is build to avoid the interlayer interactions. Lattice constants and atomic positions are fully optimized with the conjugate gradient (CG) scheme until the maximum force acting on all atoms is less than 1$\times$10$^{-3}$ eV/\AA~and the total energy was converged to 10$^{-7}$ eV. The $9\times9\times1$ and $15\times15\times1$ K-meshes generated by $\Gamma$-centered Monkhorst-Pack grid~\cite{Monkhorst1976} are used for structure optimization and self-consistent calculations. The plane-wave cutoff energy is set to be 500 eV. The phonon frequency calculations have been carried out using the finite displacement approach as implemented in the PHONOPY code\cite{Togo2015} with a $4\times4\times1$ supercell. The thermal stability is examined by performing a molecular dynamics (MD) simulations in the canonical ($NVT$) ensemble in a $4\times4\times1$ supercell at different temperatures with a Nos\'{e} thermostat.  In the calculation of Co$_3$Sn$_3$Se$_2$/MoS$_2$ heterostructure, zero damping DFT-D3 method is adopted to take interlayer van der Waals forces into account. Surface states are investigated by an effective tight-binding Hamiltonian constructed from the maximally localized Wannier functions\cite{Mostofi2014,Kong2017}. And the iterative Green function method\cite{Sancho1985} is used with the package WannierTools\cite{Wu2018}.

\section{\uppercase\expandafter{\romannumeral3}. DFT results of monolayers Co$_3$Pb$_3$S$_2$, Co$_3$Pb$_3$Se$_2$ and Co$_3$Sn$_3$Se$_2$ }
\subsection{A. Crystal structure and stability}

The crystal structure of Co$_3$X$_3$Y$_2$ (X = Sn,Pb; Y = S,Se) with the space group of $\textit{P}$\={3}$\textit{m}$1 (No. 164) is depicted in Figs.\ref{fig1}(a) and (b). Co atoms form a 2D kagome lattice with one X atom sandwiched by X and Y atoms in the center. Each primitive cell contains one formula unit. After checking the dynamical stabilities of all compounds Co$_3$X$_3$Y$_2$ (X = C,Si,Ge,Sn,Pb; Y =O,S,Se,Te,Po) by calculating their phonon spectra, we found that only Co$_3$Pb$_3$S$_2$, Co$_3$Pb$_3$Se$_2$ and Co$_3$Sn$_3$Se$_2$ monolayers are dynamically stable, because there are no imaginary phonon modes in the whole Brillouin zone as shown in Figs.\ref{fig1}(d)-(f). The calculated lattice constants for Co$_3$Pb$_3$S$_2$, Co$_3$Pb$_3$Se$_2$ and Co$_3$Sn$_3$Se$_2$ are 5.38, 5.44 and 5.32 {\AA}, respectively. Furthermore, the stabilities of these compounds are also checked by the formation energy, which is defined as $E_f = E_{Co_3X_3Y_2} - 3E_{Co} - 3E_{X} - 2E_{Y}$, where $E_{Co_3X_3Y_2}$ is the energy of Co$_3$X$_3$Y$_2$ monolayer, $E_{Co}$, $E_{X}$ and $E_{Y}$ are the energies of Co, X and Y crystals, respectively. The obtained negative values -2.31, -1.02 and -1.86 eV indicate exothermic reactions for Co$_3$Pb$_3$S$_2$, Co$_3$Pb$_3$Se$_2$ and Co$_3$Sn$_3$Se$_2$. Their thermal stabilities are then confirmed by the molecular dynamics simulations performed in a $4\times4\times1$ supercell at 300, 500 and 700K, respectively. As shown in Figs.\ref{fig1}(g)-(i), the small fluctuation of total energy indicates their thermal stabilities.

\subsection{B. Magnetic and electronic properties}

\begin{table}
\begin{center}
\caption{The calculated energy difference ($\Delta$E) between the ferromagnetic and antiferromagnetic  configurations per primitive cell, magnetic moment (M), magnetic anisotropy energy (MAE), band gap,  nearest-neighbor exchange integral (J$\lvert{S}\rvert^2$) where $S$ is the spin at Co atom, single-ion anisotropy (D$\lvert{S}\rvert^2$) , Curie temperature (T$_C$) and the maximum Kerr angle ($\theta_\text{Kerr}$) for Co$_3$Pb$_3$S$_2$, Co$_3$Pb$_3$Se$_2$ and Co$_3$Sn$_3$Se$_2$ monolayers.}\label{tab:table1}
\centering
\setlength{\tabcolsep}{1.7mm}{
\begin{tabular}{lcccc}
\hline
              &  Co$_3$Pb$_3$S$_2$       &    Co$_3$Pb$_3$Se$_2$  &    Co$_3$Sn$_3$Se$_2$  \\
\hline
$\Delta$E (meV)  &     -60.5                &          -44.8         &          -51.5   \\
M ($\mu_B$)   &      0.42                &           0.43         &           0.39   \\
MAE (meV/Co)  &      0.42                &           0.73         &           0.65   \\
Gap (meV)     &     70                   &            77          &            63    \\
J$\lvert{S}\rvert^2$ (meV)  &  6.73             &           4.98         &         5.73   \\
D$\lvert{S}\rvert^2$ (meV) &  0.42             &           0.73         &         0.65   \\
T$_c$ (K)     &     51                   &            42          &            46    \\
$\theta_\text{Kerr}$ (deg) &  2.25       &           2.83         &           1.28   \\
\hline
\end{tabular}}
\end{center}
\end{table}

\begin{figure}[!hbt]
  \centering
  \includegraphics[scale=0.85,angle=0]{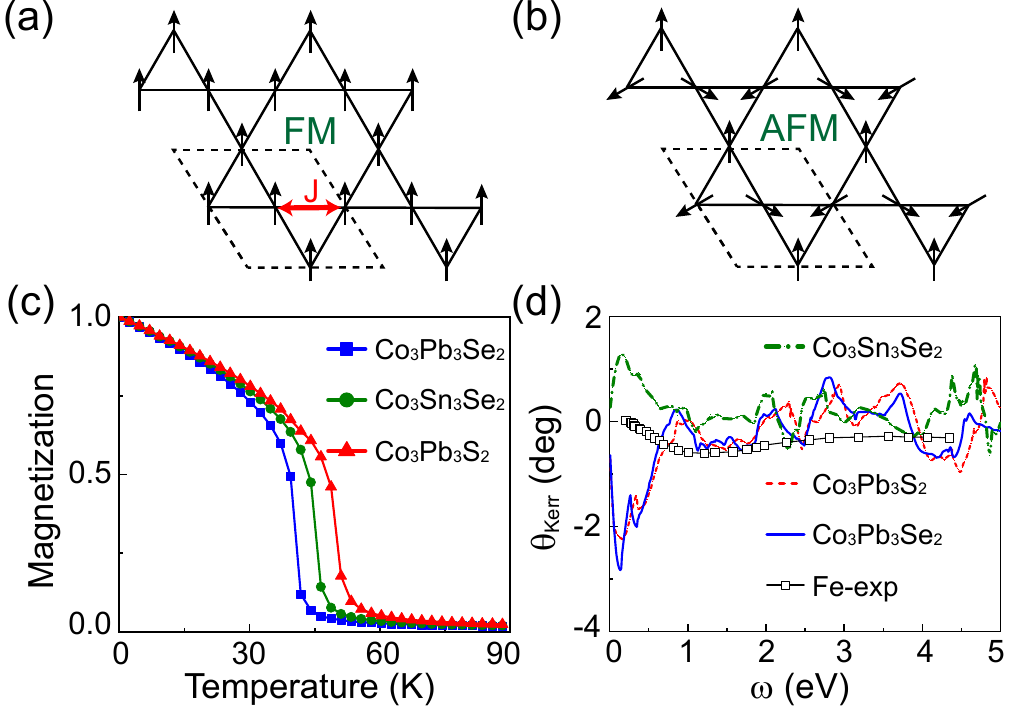}\\
  \caption{(a) Ferromagnetic (FM) and (b) antiferromagntic (AFM) spin configurations for Co atoms in the kagome lattice. For Co$_3$Pb$_3$S$_2$, Co$_3$Pb$_3$Se$_2$ and Co$_3$Sn$_3$Se$_2$ monolayers, (c) temperature dependent normalized magnetic moment, and (d) Kerr angle as a function of photon energy.}\label{fig2}
\end{figure}

In DFT calculations, the magnetic ground state of Co$_3$X$_3$Y$_2$ monolayer can be obtained by calculating the energy difference between ferromagnetic (FM) and antiferromagnetic (AFM) spin configurations($\Delta$$E$ = $E_{FM}$ - $E_{AFM}$), as shown in Figs.\ref{fig2}(a) and (b). The AFM configuration is consisting of in-plane spin polarization with angles of 120~\cite{Liu2016}. The results are listed in Table.\ref{tab:table1}, where the negative values of energy difference between the FM and AFM configurations indicate that Co$_3$Pb$_3$S$_2$, Co$_3$Pb$_3$Se$_2$ and Co$_3$Sn$_3$Se$_2$ all favor FM ground state. For the FM ground state, the magnetic anisotropy energy (MAE) which is defined as the energy difference between total energies corresponding to in-plane and out-of-plane FM configurations for these three compounds were calculated as listed in Table.\ref{tab:table1}. One may see that all these compounds prefer an out-of-plane magnetization.

The magnetism of these compounds can be described by the following Heisenberg-type Hamiltonian
\begin{equation}\label{Hamiltonian}
H_0 = -\sum_{\langle i,j\rangle}J\boldsymbol{S_i}\cdot\boldsymbol{S_j}-\sum_{\langle i\rangle}D{\boldsymbol{S_{iz}^2}},
\end{equation}
\noindent
where $J$ and $D$ are the nearest-neighbor exchange integral and single-ion anisotropy (SIA), respectively. In order to obtain J$\lvert{S}\rvert^2$ and D$\lvert{S}\rvert^2$, the energies corresponding to three different magnetic configurations: FM($\bm{m}$$\parallel$$\bm{x}$), FM($\bm{m}$$\parallel$$\bm{z}$) and AFM are expressed as
\begin{equation}
\begin{aligned}
&E_{FM(\bm{m}\parallel\bm{x})}=-6J|\boldsymbol{S}|^2+E_0, \\
&E_{FM(\bm{m}\parallel\bm{z})}=-6J|\boldsymbol{S}|^2-3D|\boldsymbol{S}|^2+E_0,\\
&E_{AFM}=3J|\boldsymbol{S}|^2+E_0,
\end{aligned}
\end{equation}
\noindent
where E$_0$ is the energy which is independent of spin configurations. The corresponding J$\lvert{S}\rvert^2$ and D$\lvert{S}\rvert^2$ are listed in Table.\ref{tab:table1}. The Monte Carlo (MC) simulations on a $80\times80\times1$ kagome lattice with periodic boundary conditions are carried out with each temperature calculations containing 10$^6$ MC steps\cite{Wolff1989}. The Curie temperatures are estimated to be 51, 42 and 46K for Co$_3$Pb$_3$S$_2$, Co$_3$Pb$_3$Se$_2$ and Co$_3$Sn$_3$Se$_2$ monolayers, respectively, as shown in Fig.~\ref{fig2}(c).

\begin{figure*}[!hbt]
  \centering
  \includegraphics[scale=0.8,angle=0]{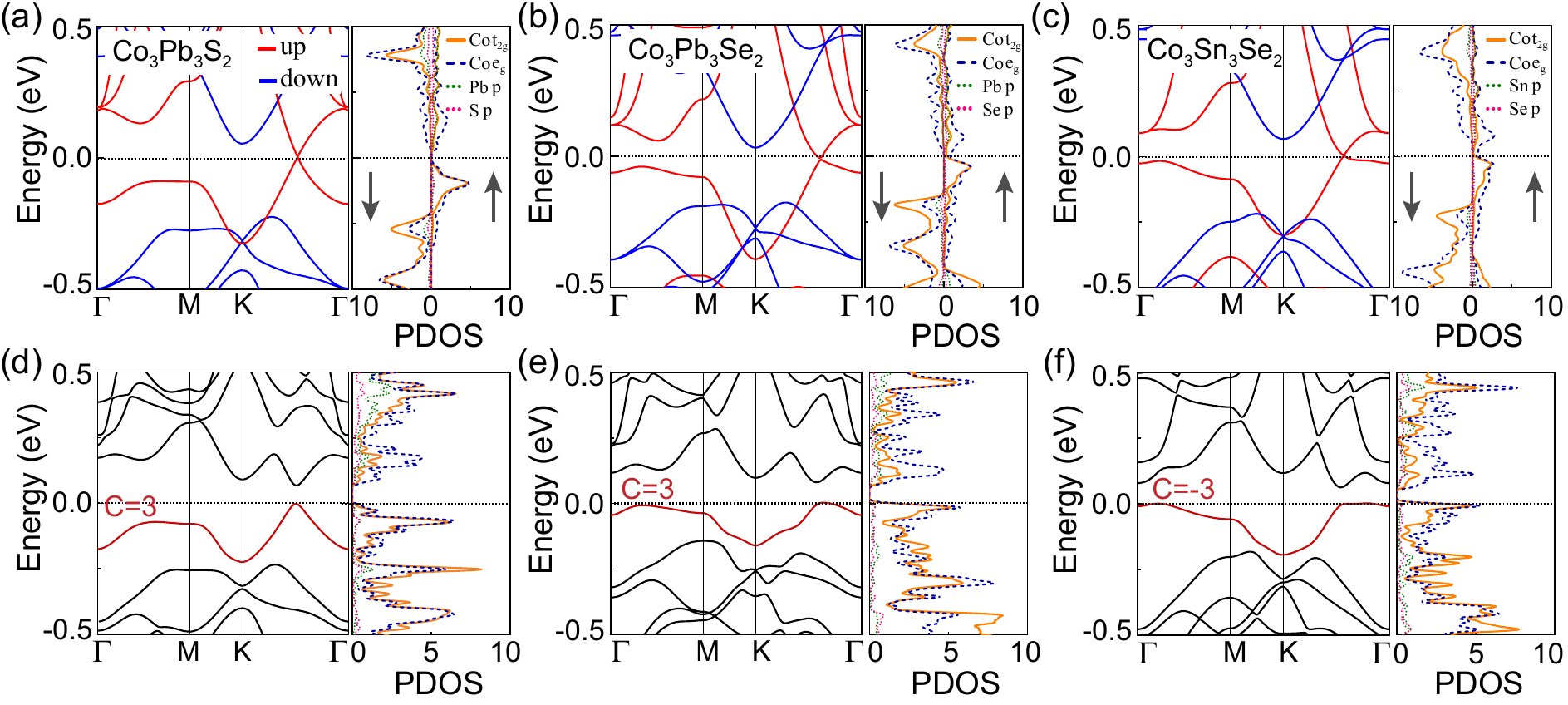}\\
  \caption{Electronic band structure and projected density of states for monolayers (a) Co$_3$Pb$_3$S$_2$, (b) Co$_3$Pb$_3$Se$_2$ and (c) Co$_3$Sn$_3$Se$_2$ without spin-orbit coupling (SOC). (d)-(f) the corresponding results with inclusion of SOC, respectively.}\label{fig3}
\end{figure*}

The magneto-optical Kerr effect is usually expected in FM materials, due to their potential applications in magneto-optical storage devices. The Kerr rotation angle can be written as
$\theta_\text{Kerr}=-\text{Re}\frac{\epsilon_\text{xy}}{(\epsilon_\text{xx}-1)\sqrt{\epsilon_\text{xx}}}$,
where $\omega$ is the photon energy, $\epsilon_\text{xx}$ and $\epsilon_\text{xy}$ are the diagonal and off-diagonal terms of the dielectric tensor $\epsilon$. And the dielectric tensor $\epsilon$ has a relationship with optical conductivity tensor $\sigma$ that could be expressed as
$\sigma(\omega)=\frac{\omega}{4{\pi}i}[\epsilon(\omega)-I]$,
where $I$ is a unit tensor. Optical conductivity tensor $\sigma$ and Kerr rotation angle can be obtained using VASP along with WANNIER90. The photon energy dependent Kerr angles are illustrated in Fig.\ref{fig2}(d). The sign of Kerr angles of Co$_3$Pb$_3$S$_2$ and Co$_3$Pb$_3$Se$_2$ differ from Co$_3$Sn$_3$Se$_2$ in the low photon energy range, due to the opposite sign of $\epsilon_{xy}$.  For these three compounds, at the photon energy of about 0.2 eV Kerr angles reach to the maximum values of about 2.9 as listed in Table.\ref{tab:table1}, which is much larger than the reported 0.8 for Fe bulk and is comparable to the Tc-based ferromagnetic semiconductors~\cite{You2020b}.

\begin{figure*}[!hbt]
  \centering
  \includegraphics[scale=0.9,angle=0]{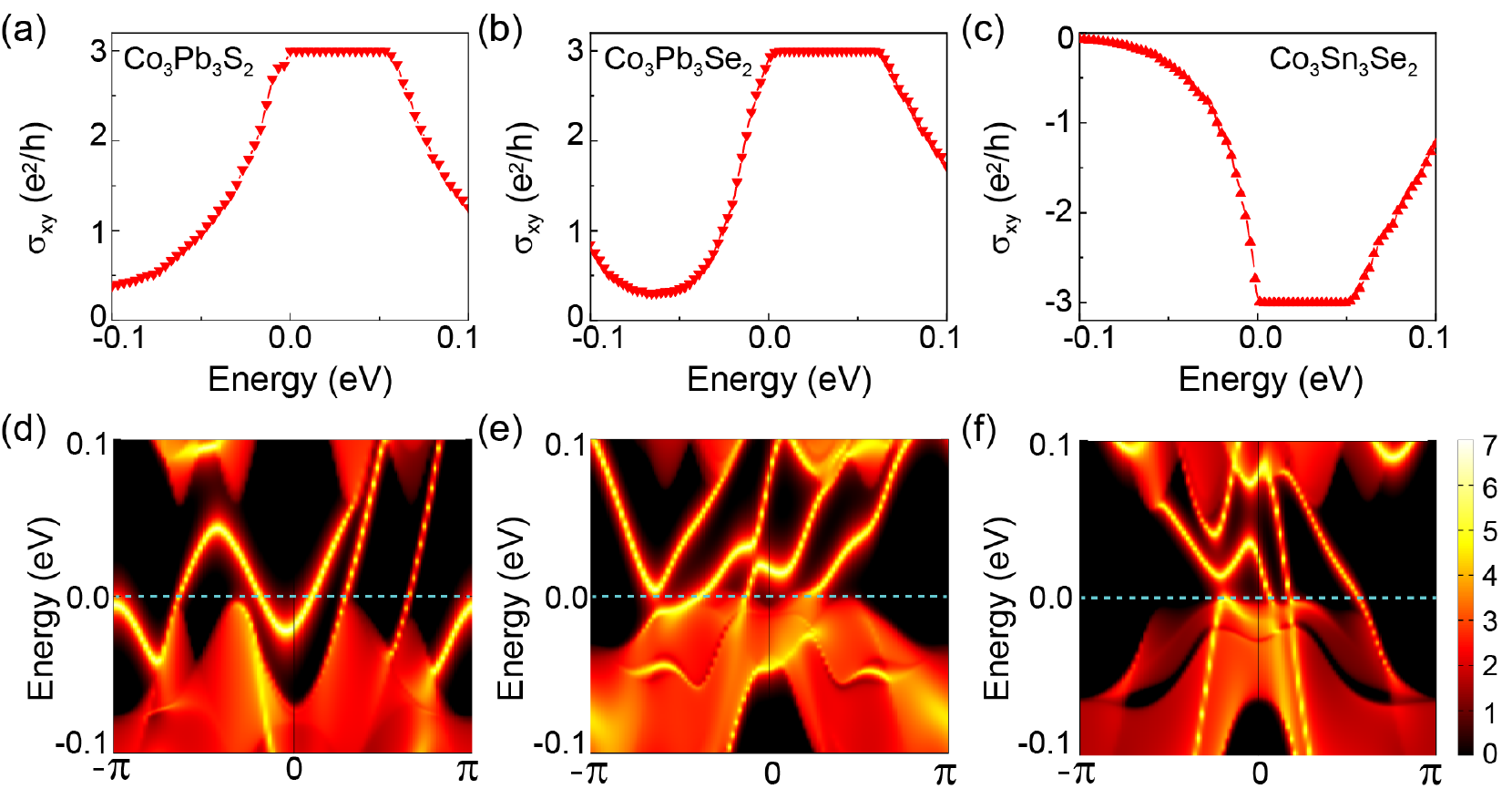}\\
  \caption{ The anomalous Hall conductance for monolayers (a) Co$_3$Pb$_3$S$_2$, (b) Co$_3$Pb$_3$Se$_2$ and (c) Co$_3$Sn$_3$Se$_2$. (d)-(f) the corresponding chiral edge states projected on the (100) surface where the bright lines showing the chiral edge states, respectively.}\label{fig4}
\end{figure*}

The spin-polarized band structures were calculated as shown in Fig.\ref{fig3}. In the absence of spin-orbit coupling (SOC), they all behavior as a Weyl half-semimetal~\cite{You2019a} with a fully spin-polarized Weyl point on the $K$$\rightarrow$$\Gamma$ path near the Fermi level. Because of the symmetries, there should be three pairs of Weyl nodes in the whole BZ. The corresponding partial density of states (PDOS) show that the density of states near the Fermi level is mainly attributed to the $t_{2g}$ and $e_g$ orbitals of Co atoms. After including SOC, the band gaps of about 70, 77 and 63 meV are opened for Co$_3$Pb$_3$S$_2$, Co$_3$Pb$_3$Se$_2$ and Co$_3$Sn$_3$Se$_2$, respectively.

\subsection{C. Topological properties}

In order to investigate their topological properties, maximally localized Wannier functions (MLWFs) implemented in the WANNIER90 package are employed to fit their DFT band structures. Nonzero Chern number $C$ is viewed as a character of topologically nontrivial band structure, and for each band, the Chern number can be obtained by integrating the Berry curvature over the first Brillouin zone. The calculated Chern number $C$ for the valence band near the Fermi level is marked in Fig.\ref{fig3}. Figs.\ref{fig4}(a)-(c) show the calculated anomalous Hall conductance as a function of the chemical potential, and the quantized charge Hall plateau of $\sigma_{xy} = Ce^2/h$ is obtained when chemical potential is within the band gap, characterizing the QAHE. Furthermore, the QAHE can also be confirmed by calculating their chiral edge states appearing within the band gap. On the basis of a recursive strategy, we construct the MLWFs using all $d$ orbitals of Co atoms and calculate their local density of the edge states as shown in Figs.\ref{fig4}(d)-(f). One can see that the bulk states are connected by three topologically nontrivial edge states. As the number of edge states cutting the Fermi level indicates the value of the Chern number, $\lvert$$C$$\rvert$ = 3 is further verified. The internal of Berry curvature in a region around one original Weyl point gives a Chern number of 1/2. There are six Weyl points related by $C_3$ and inversion symmetry that should have the same Chern number of 1/2. Thus we can obtain the Chern number 3.

\subsection{\uppercase\expandafter{\romannumeral4}. STRAIN EFFECT AND Co$_3$Sn$_3$Se$_2$/MoS$_2$ HETEROSTRUCTURE}

\begin{figure}[!hbt]
  \centering
  \includegraphics[scale=0.7,angle=0]{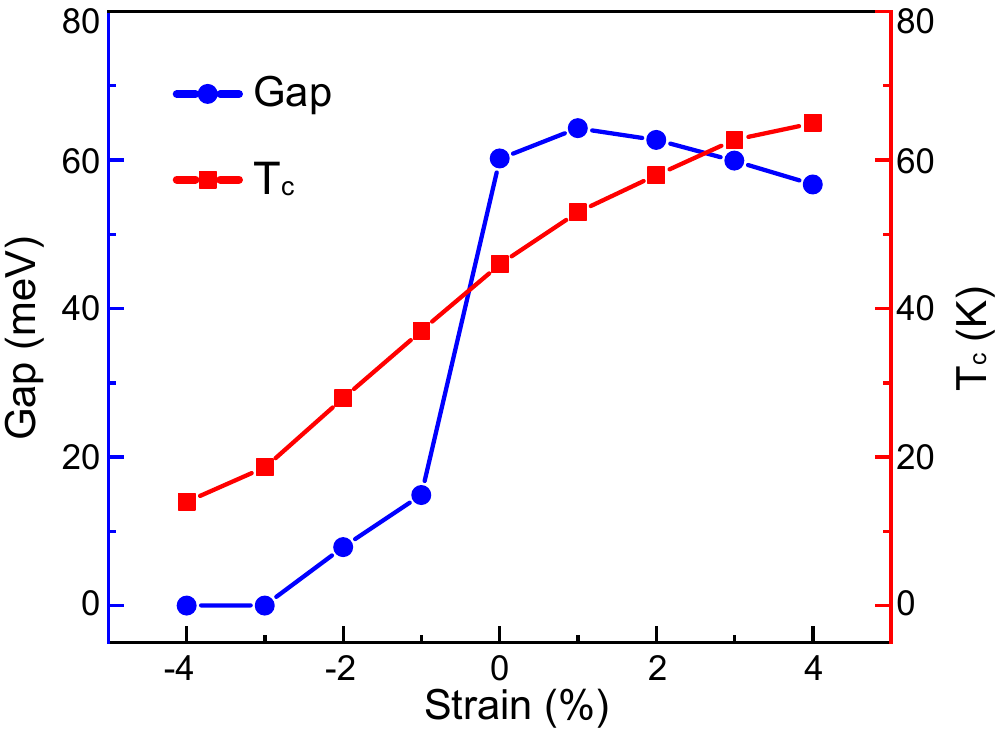}\\
  \caption{Band gap and Curie temperature as functions of applied biaxial strain of monolayer Co$_3$Sn$_3$Se$_2$.}\label{fig5}
\end{figure}

To investigate the effect of strain on Co$_3$X$_3$Y$_2$ monolayers, as shown in Fig.\ref{fig5}, we plot the band gap and Curie temperature $T_C$ of Co$_3$Sn$_3$Se$_2$ monolayer as the function of biaxial strain which is defined as $\varepsilon = (a-a_0)/a_0$, where $a$ and $a_0$ are the strained and equilibrium lattice parameters, respectively.
With the increase of compressed strain, the valence band maximum (VBM) value at $\Gamma$ point is lifted, while the conduction band minimum (CBM) value drops at K point, leading to the gradual closure of band gap. The VBM at $\Gamma$ point and CBM at K point move in the opposite directions relative to the Fermi level with the tensile strain, and thus the band gap keeps about 60 meV. Meanwhile, the applied tensile strain will enhance the exchange-coupling parameter $J$ and the corresponding Curie temperature from 46 K to 65 K. The strain effect on band gap and $T_c$ for Co$_3$Pb$_3$S$_2$ and Co$_3$Pb$_3$Se$_2$ has also been studied as shown in Fig.S2, where similar behavior is obtained.

\begin{figure}[!hbt]
  \centering
  \includegraphics[scale=0.8,angle=0]{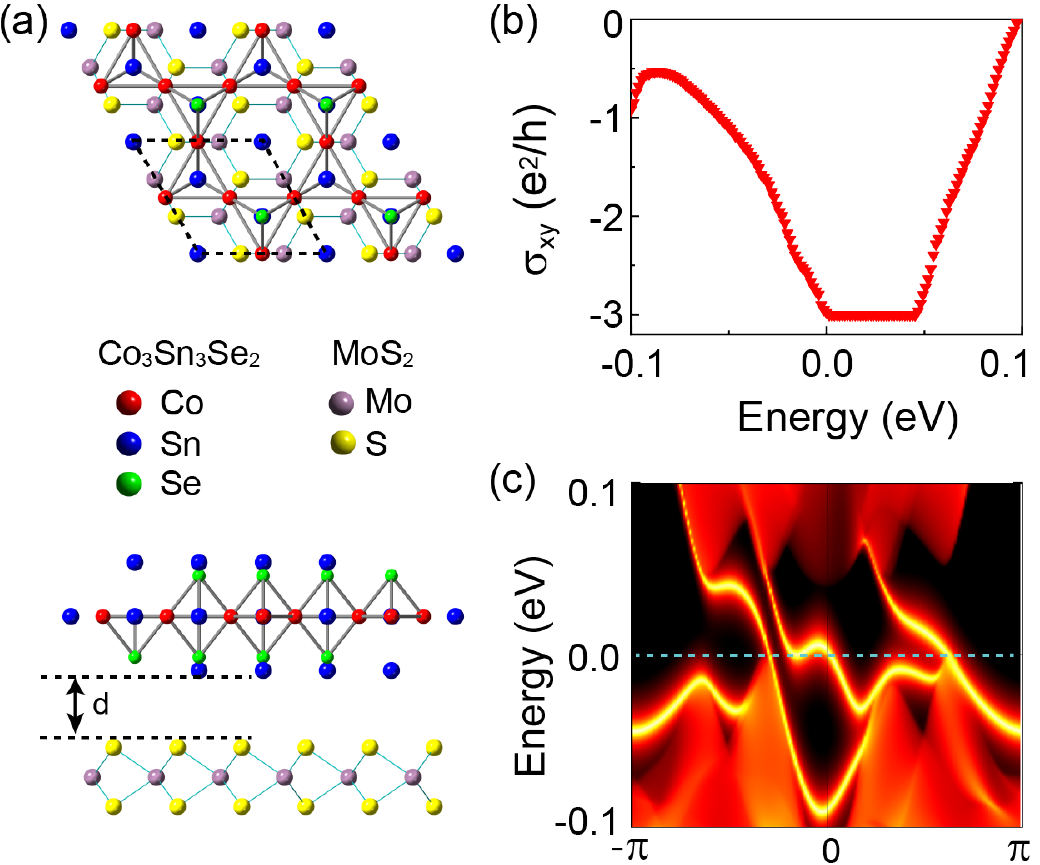}\\
  \caption{For the heterostructure Co$_3$Sn$_3$Se$_2$/MoS$_2$, (a) top and side views of crystal structure, (b) anomalous Hall conductance, and (c) chiral edge states.}\label{fig6}
\end{figure}

To investigate the effect of the substrate on electronic properties of Co$_3$X$_3$Y$_2$ monolayers, we construct a heterostructure Co$_3$Sn$_3$Se$_2$/MoS$_2$ as shown in Fig.~\ref{fig6}(a). The heterostructure is consisting of $1\times1$ unit cell of Co$_3$Sn$_3$Se$_2$ and $\sqrt{3}\times\sqrt{3}$ unit cell of MoS$_2$. The lattice mismatch at interface is only 2\%. By the first-principles calculations with the vdW interaction, the optimized lattice constant and equilibrium interlayer distance $d$ for the heterostructure is about 5.42{\AA} and 2.76{\AA}, respectively. The $T_C$ is enhanced up to 60 K, with the band gap keeps about 60 meV. The calculated AHC and edge states as shown in Figs.\ref{fig6}(b) and (c) indicate that the topological property of Co$_3$Sn$_3$Se$_2$ preserves.

\subsection{\uppercase\expandafter{\romannumeral5}. PROPERTIES of Co$_6$Pb$_5$S$_4$, Co$_6$Pb$_5$Se$_4$ AND Co$_6$Sn$_5$Se$_4$ BILAYERS}

\begin{figure}[!hbt]
  \centering
  \includegraphics[scale=0.8,angle=0]{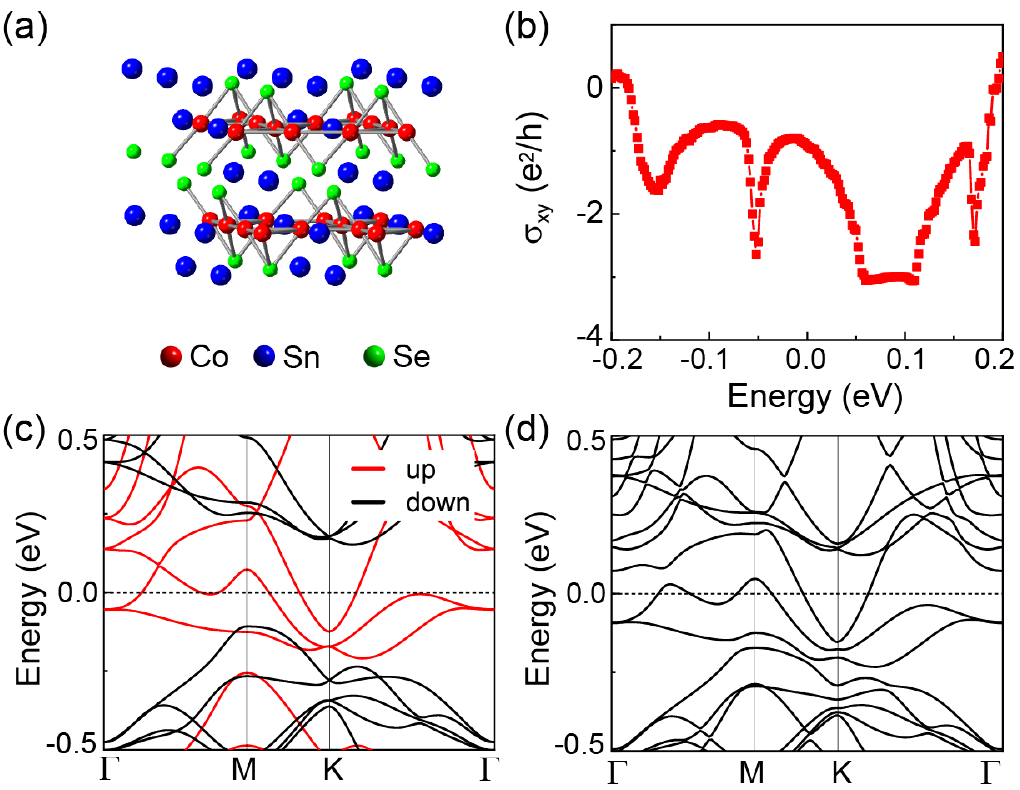}\\
  \caption{For bilayer compound Co$_6$Sn$_5$Se$_4$, (a) crystal structure, (b) band structure without spin-orbit coupling (SOC), (c) band structure with SOC, and (d) anomalous Hall conductance. }\label{fig7}
\end{figure}

To study the layer dependent topological properties, the stabilities of the bilayer compounds Co$_6$Pb$_5$S$_4$, Co$_6$Pb$_5$Se$_4$ and Co$_6$Sn$_5$Se$_4$ are first confirmed by calculating the phonon spectra, where no imaginary frequency was observed as shown in Fig.S3. For Co$_6$Sn$_5$Se$_4$, we have noticed that a Weyl node appears at the $\Gamma\rightarrow$ M path near the Fermi level as illustrated in Fig.\ref{fig7}(b). Taking the SOC into account, a local band gap of about 51 meV at the (original) Weyl point is opened. Although there is no global gap, it is interesting to find a relatively flat plateau of AHC corresponding to $\lvert$$C$$\rvert$ = 3 near the Fermi level.

\section{\uppercase\expandafter{\romannumeral6}. CONCLUSION}

By using first principles calculations, we have systematically investigated two-dimensional kagome ferromagnets Co$_3$Pb$_3$S$_2$, Co$_3$Pb$_3$Se$_2$ and Co$_3$Sn$_3$Se$_2$ monolayers, which can realize the high-Chern-number ($\lvert$$C$$\rvert$ = 3) QAHE with a large band gap. For Co$_3$Pb$_3$S$_2$, Co$_3$Pb$_3$Se$_2$ and Co$_3$Sn$_3$Se$_2$ monolayers, the band gap of 70, 77 and 63 meV and $T_C$ of 51, 42, and 46K are obtained, respectively. By constructing a heterostructure Co$_3$Sn$_3$Se$_2$/MoS$_2$, the $T_C$ can be enhanced to 60 K and the band gap keeps about 60 meV due to the tensile strain of 2\% at the interface. For the bilayer compound Co$_6$Sn$_5$Se$_4$, it becomes a half-metal, with a relatively flat plateau in its anomalous Hall conductivity corresponding to $\lvert$$C$$\rvert$ = 3 near the Fermi level. Our results provide new topologically nontrivial systems of kagome ferromagnetic monolayers and heterostructrues with high $\lvert$$C$$\rvert$ = 3  and large band gap QAHE, which are helpful for us to deepen understanding on the topological states in ferromagnets with kagome lattices.

\section{Acknowledgements}
This work is supported in part by the National Key R$\&$D Program of China (Grant No. 2018YFA0305800), the Strategic Priority Research Program of the Chinese Academy of Sciences (Grant No. XDB28000000), the
National Natural Science Foundation of China (Grant No.11834014), and Beijing Municipal Science and Technology Commission (Grant No. Z191100007219013). B.G. is also supported by the National Natural Science Foundation of China (Grants No. Y81Z01A1A9 and No. 12074378), the Chinese Academy of Sciences (Grants No. Y929013EA2 and No. E0EG4301X2), the University of Chinese Academy of Sciences (Grant
No. 110200M208), the Strategic Priority Research Program of Chinese Academy of Sciences (Grant No. XDB33000000), and the Beijing Natural Science Foundation (Grant No. Z190011).

\bibliographystyle{apsrev4-1}
\bibliography{ref}

\end{document}